\documentclass[journal=aamick,manuscript=article]{achemso}

\usepackage[version=3]{mhchem} 
\usepackage[T1]{fontenc}       
\usepackage[utf8]{inputenc}
\usepackage[english]{babel}
\usepackage{amsmath,amsfonts,amsthm,amssymb}
\usepackage{mathtools}
\usepackage{graphicx}
\usepackage{xcolor}
\usepackage[pdftex]{hyperref}
\usepackage{bm}
\usepackage[inline]{enumitem}
\usepackage{array}
\usepackage{gensymb}
\usepackage[font=footnotesize,labelfont=default]{caption}
\usepackage{verbatim}
\usepackage{mathrsfs}
\usepackage{siunitx}
\usepackage{chemformula}

\renewcommand{\eqref}[1]{Eq.~(\ref{eq:#1})}

\newcommand{\imag}{\mathrm{i}}

\author{Sergejs~Boroviks}
\affiliation[]{Center for Nano Optics, University of Southern Denmark, Campusvej 55, DK-5230~Odense~M, Denmark}

\author{Christian Wolff}
\affiliation[]{Center for Nano Optics, University of Southern Denmark, Campusvej 55, DK-5230~Odense~M, Denmark}

\email{cwo@mci.sdu.dk}

\author{Jes Linnet}
\affiliation[]{Center for Nano Optics, University of Southern Denmark, Campusvej 55, DK-5230~Odense~M, Denmark}

\author{Yuanqing Yang}
\affiliation[]{Center for Nano Optics, University of Southern Denmark, Campusvej 55, DK-5230~Odense~M, Denmark}

\author{Francesco~Todisco}
\affiliation[]{Center for Nano Optics, University of Southern Denmark, Campusvej 55, DK-5230~Odense~M, Denmark}

\author{Alexander~S.~Roberts}
\affiliation[]{Center for Nano Optics, University of Southern Denmark, Campusvej 55, DK-5230~Odense~M, Denmark}

\author{Sergey~I.~Bozhevolnyi}
\affiliation[]{Center for Nano Optics, University of Southern Denmark, Campusvej 55, DK-5230~Odense~M, Denmark}
\alsoaffiliation[]{Danish Institute for Advanced Study, University of Southern Denmark, Campusvej 55, DK-5230~Odense~M, Denmark}

\author{Bert Hecht}
\affiliation[]{Nano-Optics and Biophotonics Group, Experimentelle Physik 5, Physikalisches Institut, Wilhelm-Conrad-R{\"o}ntgen-Center for Complex Material Systems, Universit{\"a}t W{\"u}rzburg, W{\"u}rzburg, Germany}
\author{N.~Asger~Mortensen}
\affiliation[]{Center for Nano Optics, University of Southern Denmark, Campusvej 55, DK-5230~Odense~M, Denmark}
\alsoaffiliation[]{Danish Institute for Advanced Study, University of Southern Denmark, Campusvej 55, DK-5230~Odense~M, Denmark}

\email{asger@mailaps.org}

\title{Interference in edge-scattering from monocrystalline gold flakes}


\begin{document}

\begin{abstract}
  We observe strongly dissimilar scattering from two types of edges in hexagonal quasi-monocrystalline gold flakes with thicknesses around 1 micron.
  We identify as the origin the interference between a direct, quasi-specular scattering and an indirect scattering process involving an intermediate surface-plasmon state.
  The dissimilarity between the two types of edges is a direct consequence of the three-fold symmetry around the [111]-axis and the intrinsic chirality of a face-centered cubic lattice. We propose that this effect can be used to estimate flake thickness, crystal morphology, and surface contamination.
\end{abstract}

\section{Introduction}

Historically, the field of plasmonics\cite{Fernandez-Dominguez:2017} has explored the interaction of light with the free electron gas, with a predominant attention to amorphous and polycrystalline noble metal nanostructures and thin films,\cite{McPeak:2015} while less attention has been devoted to plasmons supported by monocrystalline materials. More recently, chemically synthesized monocrystalline gold flakes have been receiving increasing attention within the plasmonic community. In many aspects, such colloidal gold nanoparticles show superior plasmonic properties, as compared to evaporated polycrystalline films.\cite{Huang:2010:ncomm, Wu:2015:CrystResTechnol, Hoffmann:2016:Nanoscale,Krauss:2018:CrystGrowthDes} Atomic flatness and well-defined crystal structure offer larger plasmon propagation lengths and sharper resonances due to lower Ohmic losses and reduced surface scattering.\cite{Mejard:2017:OME} These favorable properties have been utilized in the design and fabrication of various plasmonic devices, such as nano-circuits,\cite{Huang:2010:ncomm, Dai:2014:NanoLett} nano-antennas,\cite{Prangsma:2012:NanoLett, Chen:2014:ACSNano, Kern:2015:NatPhotonics} tapers,\cite{Schmidt:2012} and plasmon billiards.\cite{Spektor:2017,Frank:2017}
However, flat metal crystals are rarely true single crystals, but rather twins joined at pairs of stacking faults.\cite{Lofton:2005:AdvFuncMat}
This is no coincidence, because the strong lateral growth involving [100]-facets requires the presence of at least 2 stacking faults within the seed.\cite{Krauss:2018:CrystGrowthDes}
These defects play an important role in the crystal growth \cite{Hoffmann:2016:Nanoscale} and might exhibit interesting plasmonic phenomena of electronic 2D states.
This along with the well defined material properties of single crystals render them excellent candidates for the observation of quantum effects\cite{Ding:2017:PhysRevB,Cuche:2017:PhysRevB} or of anisotropic nonlinear or nonlocal response e.g. due to the deviation of the Fermi surface from a perfect sphere.
This is especially true for sub-micron particles as quantum corrections to the classical electrodynamics manifest increasingly when approaching the subwavelength scale and reaching out for atomic dimensions.\cite{Bozhevolnyi:2017,Fernandez-Dominguez:2018} 

Differences between the material properties of mono- and polycrystalline metals are both important for applications as well as interesting in their own right.
Moreover, the highly ordered atomic structure of single crystals is also reflected in their geometry with very well defined angles, atomically flat surfaces and sharp edges.
The quality of these features is well beyond what is currently achievable with state-of-the-art nano patterning of polycrystalline films, \cite{Xiong:2014} and they provide valuable material for the basic research of nanoplasmonics in finite-size metal geometries e.g. on the propagation of surface-plasmon polaritons (SPP) along the edges \cite{Si:2015:SciRep,Major:2013:J.Phys.Chem.C} of gold crystals.
Several examples for such studies have been conducted for particles that are small compared to the wavelength of light.\cite{Feng:2015:AdvOptMat, Viarbitskaya:2013:AppPhysLett, Cuche:2015:SciRep, Viarbitskaya:2013:NatMaterials}

\begin{figure}
  \begin{center}
    \includegraphics[width=\linewidth]{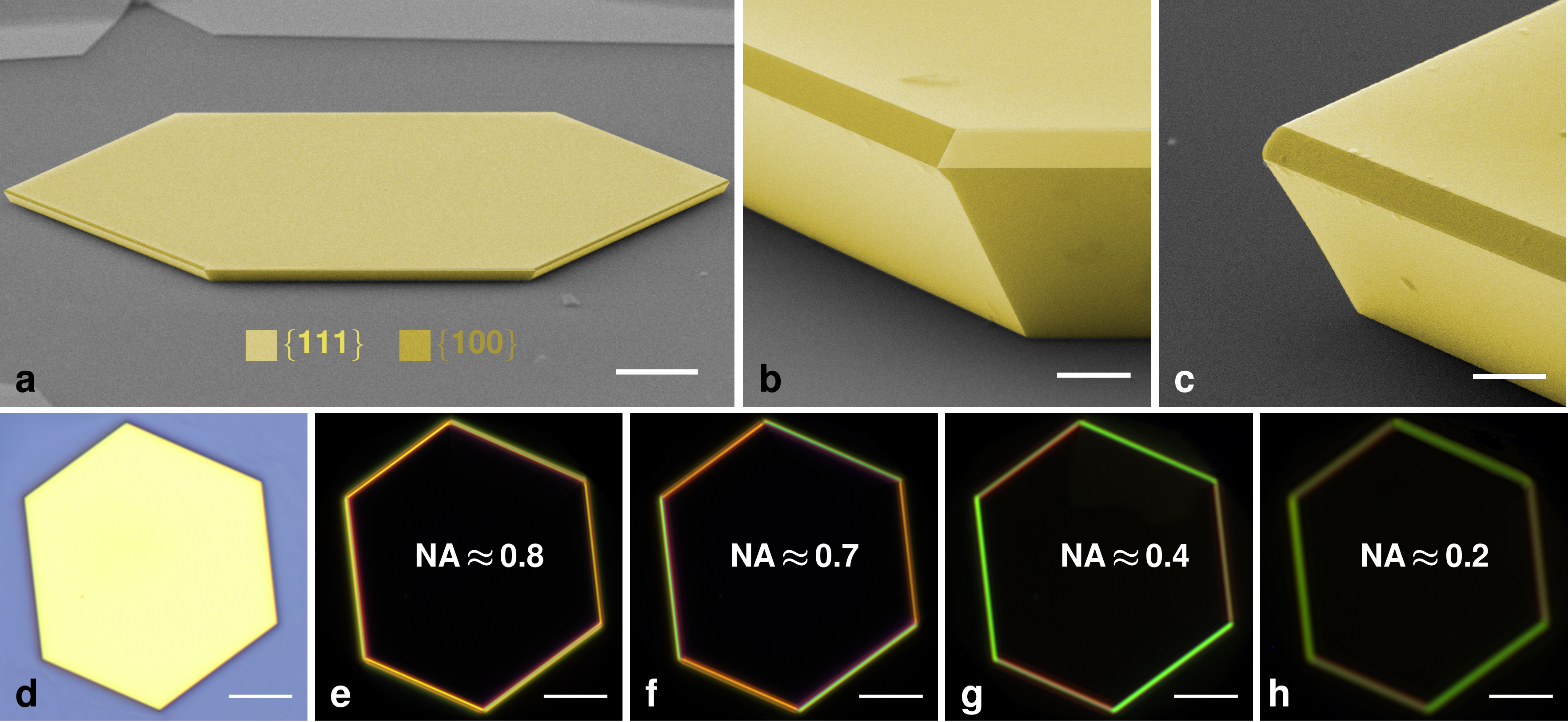}
  \end{center}\linespread{1.3}
  \caption{(a) SEM image of the Au monocrystalline flake (75$^\circ$ tilted view, scale bar: \SI{10}{\micro \meter}) (b) and (c) close-up high resolution SEM image of the two corners of the flake (75$^\circ$ tilted view, scale bars: \SI{500}{\nano \meter}); Artificial coloration is used to highlight different crystallographic planes of the facets: \{111\} (light yellow) and \{100\} (dark yellow); (d) bright-field optical image of the flake and (e-h) dark-field optical images of the flake captured with 4 different NA's (indicated in the images, scale bars: \SI{10}{\micro \meter})} 
  \label{fig:overview}
\end{figure}

Crystal-related morphologic features are of course by no means restricted to subwavelength-sized particles, but also appear in fairly large objects such as the gold flakes studied in this paper with lateral dimensions greater than \SI{10}{\micro\meter} and thicknesses around \SI{1}{\micro\meter}. 
One such non-trivial feature is the fact that the edges of our nearly hexagonal flakes are asymmetrically tapered and that two different types of such edge terminations alternate around the flake.
As a result, each edge is dissimilar to both adjacent edges and the opposing one.
This lack of symmetry with respect to mirroring and 180$^\circ$ rotation can be seen in high-resolution scanning-electron microscope (SEM) images (Figure~\ref{fig:overview}a-c). 
Yet quite often, this detail is ignored and the edges are simply approximated as rectangular truncations\cite{Si:2015:SciRep,Major:2013:J.Phys.Chem.C}.
However, it reflects the fact that the face-centered cubic (FCC) gold lattice is symmetric with respect to neither a 180$^\circ$-rotation nor a mirror operation through the [111]-axis.
It is therefore a large-scale manifestation of the atomic order and can lead to a significant difference in the optical far-field properties as we show here.

In this work we report on a distinct difference in the scattering of visible light from the two types of edge terminations of colloidally grown gold flakes with thicknesses around one micron. 
Even though in an optical bright-field (BF) our flake looks perfectly hexagonal (Figure \ref{fig:overview}d) and seemingly exhibits six fold symmetry, difference in scattering appears as differently colored edges when observed in an optical dark-field (DF) microscope under low numerical aperture (NA) collection conditions (see Figure~\ref{fig:overview}e-h). 
We conclude that this is the far-field manifestation of the fact that opposing flake edges meet the substrate at different angles, which in turn is a macroscopic consequence of the inherent chirality along the [111]-axis of the FCC lattice.

\section{Results and discussion}

\begin{figure}
\begin{center}
\includegraphics[width=0.6\columnwidth]{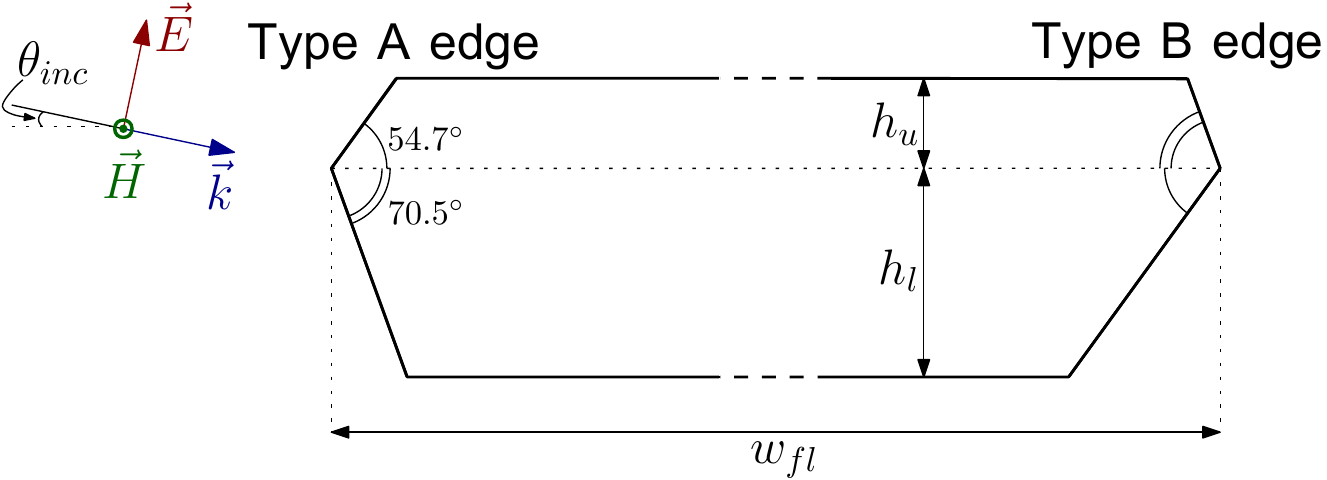}
\end{center}\linespread{1.3}
\caption{Schematic drawing of the flake's cross-section (cut trough two opposite edges) with indicated geometrical parameters.} \label{fig:schem}
\end{figure}

We grew quasi-monocrystalline gold flakes on a silicon wafer substrates from chloroauric acid in the modified Brust--Shiffrin method\cite{Brust:1994:JChemSoc} involving thermolysis of the precursor instead of chemical reduction (see Methods section for further details). This is known to yield flakes, which are large \cite{Radha:2010:CrystGrowthDes} and feature hexagonal, triangular and truncated triangular shapes with high aspect ratios up to 100. 
We obtain lateral sizes of up to 100 microns with thicknesses between several dozen nanometers and few microns. 
From the symmetries of the FCC gold lattice, one would expect the crystals to be bounded by facets of \{111\}- and \{100\}-type with threefold and fourfold symmetries, respectively.
Therefore, it seems safe to assume the large top and bottom faces to be of \{111\}-type. \cite{Hoffmann:2016:Nanoscale} As a result, each of the three or six edges of the flakes is composed of one \{111\}- and one \{100\}-type facet, which meet at an angle of $\approx 125.3^\circ$
and meet the main faces at angles of $\alpha=\arccos(1/\sqrt{3})\approx 54.7^\circ$ and $\beta=\arccos(1/3)\approx70.5^\circ$, respectively (see SEM images with artificial colorization indicating crystal planes of the facets in Figure~\ref{fig:overview}~a-c).
Consequently, the cross sections of the flakes consist of an upper and a lower trapezoid with different heights $h_u$ and $h_l$, but fixed angles, as illustrated in Figure~\ref{fig:schem}.

The strong imbalance in the size of the top and bottom \{111\}-planes over all others is due to stacking faults (more precisely multiple twin planes), which form in the early stage of the crystal growth. Such defects commonly appear in the metals with FCC crystal structure, especially in gold, as they have some of the lowest defect energies \cite{Xia:2009:AngewChemIntEd}. They lead to quite different growth rates along different crystal axes and thus cause high aspect ratio of the crystals. Therefore, such thin flakes are not strictly speaking monocrystalline, as commonly referred to, but twins.

It is noteworthy that the (111)-plane of the FCC-lattice has only three-fold symmetry even though such a crystal facet might be perfectly hexagonal, as one shown in Figure \ref{fig:overview}d. 
This is the result of the chirality of the FCC-lattice along the [111]-axis, which is due to the existence of two different stacking patterns.
Macroscopically, it manifests in the aforementioned two types of flake edges, which appear with threefold symmetry.
In the following, we refer to edges where the side facet touching the substrate is of (111)-type or (100)-type as type-A or type-B edges, respectively.
As it turns out, this three-fold symmetry and therefore the lattice chirality can be directly observed with an optical microscope. 

We noticed that flakes which look almost perfectly hexagonal in the optical bright-field and high-NA dark-field (Figure~\ref{fig:overview}d,e) exhibit very different scattering spectra from the two types of edges. This is visible to the naked eye under dark-field conditions and becomes more prominent with decreasing collection NA as illustrated in Figure~\ref{fig:overview}f-h. Images were acquired using the same objective lens and the same illumination conditions (light from DF condenser impinging on the sample at angle $\theta_{\text{inc}}\approx 12^\circ$, as shown in Figure \ref{fig:schem}), and filtering the collected light in the Fourier plane, as described in details in the Methods section.

\begin{figure}
\begin{center}
\includegraphics[width=0.8\columnwidth]{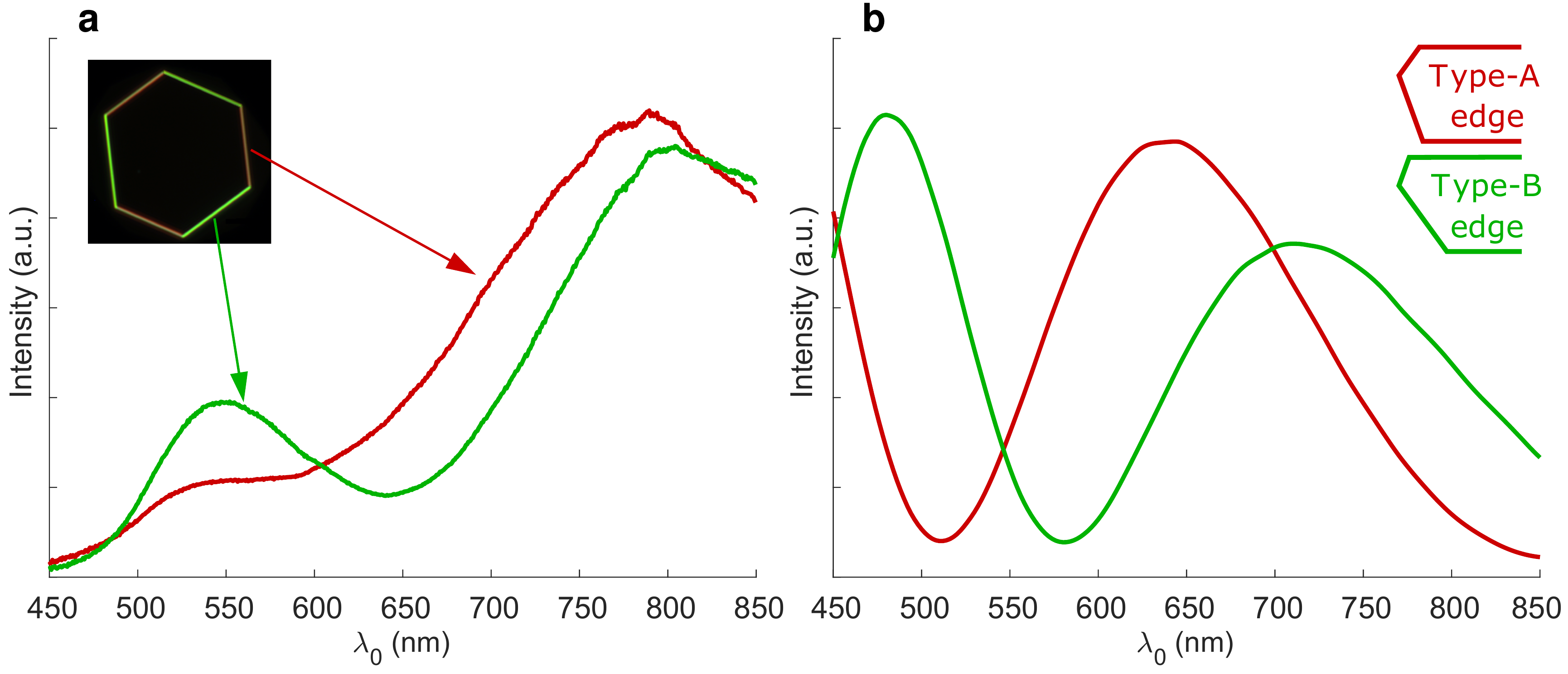}
\end{center}\linespread{1.3}
\caption{Experimental (a) and simulated (b) dark field spectra of the two types of edges of the flake with nominal dimensions $h_u=130$\,nm and $h_l=580$\,nm. Both experimental and simulated spectra are acquired with NA=0.4}\label{fig:spectra_comparison}
\end{figure}

We find behavior that is qualitatively similar to that depicted here, yet with different shades of red, yellow and green commonly for flakes with thicknesses around 1 micron, so it appears to be a geometry-related effect.
In order to further understand the underlying process, we first performed 2D finite-element calculations in p-polarization.
This is sufficient for qualitative results, because we observed experimentally that with polarized illumination only the edges perpendicular to the incident polarization appear in dark-field and that the scattered light is p-polarized itself.
We then post-processed the data with a far-field filtering procedure that mimics the effect of a low-NA objective (see Methods section for further details).
Figure~\ref{fig:spectra_comparison} shows experimental spectra acquired with NA$\approx$0.4, which provides good contrast while keeping sufficiently strong signal, for two adjacent edges of the flake shown in Figure~\ref{fig:overview}.
Alongside is shown a numerical spectrum calculated using nominal dimensions ($h_u=130$\,nm and $h_l=580$\,nm, as measured on the real flake), idealized illumination conditions and the permittivity of monocrystalline gold given by interpolated experimental data from Olmon\cite{Olmon:nAu:2012}. 
Although we clearly do not obtain quantitative agreement, the experimental
and numerical results show the same qualitative features: a peak for the type-A
edge surrounded by troughs and a similar behavior for the type-B edge 
red-shifted by some 100\,nm. 
We postpone the discussion of possible origins for this mismatch and first focus on the physical mechanism.
However, we do emphasize at this point that the spectra of different flakes differ significantly in the positions of peaks and troughs, but always show the general features of an interference pattern.

The physical origin of the different scattering spectra of type-A and type-B
edges is the interference between a surface-plasmon wave and free-space 
propagation of light, which we concluded from a careful analysis of the numerical
simulations, especially for varying values of the flake thickness parameters 
$h_u$ and $h_l$.
We observe that light scattered directly into a low-NA objective stems predominantly
from the upper facet (characterized by $h_u$), while the direct scattering
from the lower facet is directed
predominantly downwards, as one would expect 
naively from geometric optics.
Varying the parameter $h_u$ in our calculations caused broad-band changes to the scattering efficiency while leaving the beating period virtually unchanged. The upper edge appears to act as an effective point-like dipole. This holds true in simulations even if $h_u \approx h_l$. In contrast, varying $h_l$ mainly changed the beating period and strongly shifted the peaks, so the resonances are linked to the length of the lower facet.
Standing waves on that facet were considered, but did not match the observed beating period.
Light reflected from the substrate also does not enter the picture, because the illumination both in the experiment and the simulation is essentially at the Brewster angle for air/silicon interface.

\begin{figure}
\begin{center}
\includegraphics[width=0.8\columnwidth]{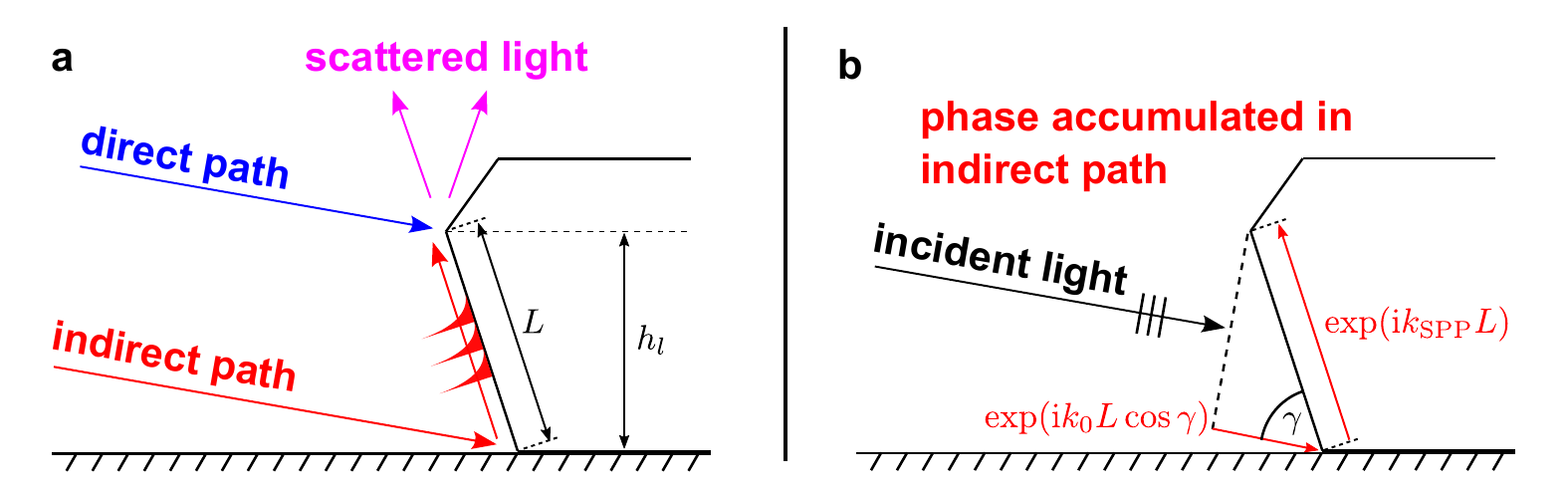}
\end{center}\linespread{1.3}
\caption{Schematic of the enhanced-scattering mechanism.
The dark-field illumination has two possible paths to scatter upwards into the objective: A direct path is the quasi-specular reflection at the upper facet of the edge.
An additional indirect path is by coupling to a surface wave at the lowest corner, traveling along the lower facet and coupling out at the upper facet.
The indirect path is delayed with respect to the direct path by a phase that depends on thickness of the lower flake part and on the angle included by the direction of incidence and the lower face.
This leads to a distinct interference effect in the far field.
  }\label{fig:principle_schematic}
\end{figure}

This leads to the following explanation: the light emanating from the upper facet has two main contributions:
Firstly, the scattered light is coming from direct illumination of the upper facet (blue arrow in Figure~\ref{fig:principle_schematic}a).
However, there is also a second, indirect path illustrated by red arrows in
Figure~\ref{fig:principle_schematic}.
Like every metallic surface discontinuity, the lower corner of the crystal couples incident light to SPP.
The resulting SPP travels up the facet, where it couples out to the far field and contributes to the scattered radiation.
Depending on the phase accumulated in this process, the direct and indirect
radiation interfere constructively or destructively in the far field, leading
to a scattering spectrum that depends sensitively on $h_l$, the edge type,
the incident angle of the DF illumination, the collection NA and the SPP dispersion 
relation. 
The in-coupling of the secondary path is significantly enhanced by the presence of a high-index substrate --- with low-index substrates the effect is much less prominent, as we observed both experimentally and in simulations. 

The phase difference between the direct and the indirect path is sketched in
Figure~\ref{fig:principle_schematic}b.
The sample is illuminated by the DF objective under a grazing angle 
$\theta_\text{inc} \approx 12^\circ$, which means that the incident light 
impinges the lower flake facet at an angle 
$\gamma = \beta - \theta_\text{inc}$ or $\gamma = \alpha - \theta_\text{inc}$ 
for type-A and type-B edges, respectively.
As a result, the upper and the lower corners of the edge are excited with a phase difference
$\Phi_1 = \exp(\imag k_0 L \cos \gamma)$, where $k_0$ is the vacuum wave number
of the light and $L$ is the length of the facet, i.e. 
$L = h_l / \sin\beta$ or $L = h_l / \sin\alpha$ for type-A and type-B edges,
respectively.
Afterwards, the indirect path gains an additional phase 
$\Phi_2 = \exp(\imag k_\text{SPP} L)$ while traveling along the facet, where
$k_\text{SPP}$ is the wave number of the SPP.
Thus, the total phase difference between the direct and the indirect path (excluding 
further phase contributions from additional processes such as standing waves 
on the lower facet) is
\begin{align}
  \Delta\Phi = \exp[\imag (k_\text{SPP} + k_0 \cos \gamma) L],\label{eq:1}
\end{align}
and we expect that the scattering spectra of either edge type depend 
periodically on the parameter $h_l$ with a periodicity of
\begin{align}
  \Delta h_l^{(\text{A})} = & 
  \frac{2\pi\sin \beta }{k_\text{SPP} + k_0 \cos(\theta_\text{inc} - \beta)} \label{eq:Delta}
\end{align}
for the type-A edge and with $\alpha$ instead of $\beta$ for the type-B
edge.
Using the permittivity $\varepsilon_\text{Au} = -16 + 1.1\imag$ for 
monocrystalline gold~\cite{Olmon:nAu:2012} at a vacuum wavelength of 
$\lambda_0 = 700\,\text{nm}$ as an example, and assuming 
$\theta_\text{inc} = 12^\circ$, we find:
\begin{align}
  \Delta h_l^{(\text{A})} \approx 420\,\text{nm} \quad \text{and} \quad
  \Delta h_l^{(\text{B})} \approx 320\,\text{nm}.
\end{align}
This means that we expect the low-NA spectra observed at type-A or type-B edges to repeat whenever the thickness of the lower part of the flake is increased by $420\,\text{nm}$ or $320\,\text{nm}$, respectively.

\begin{figure}
\begin{center}
\includegraphics[width=0.7\columnwidth]{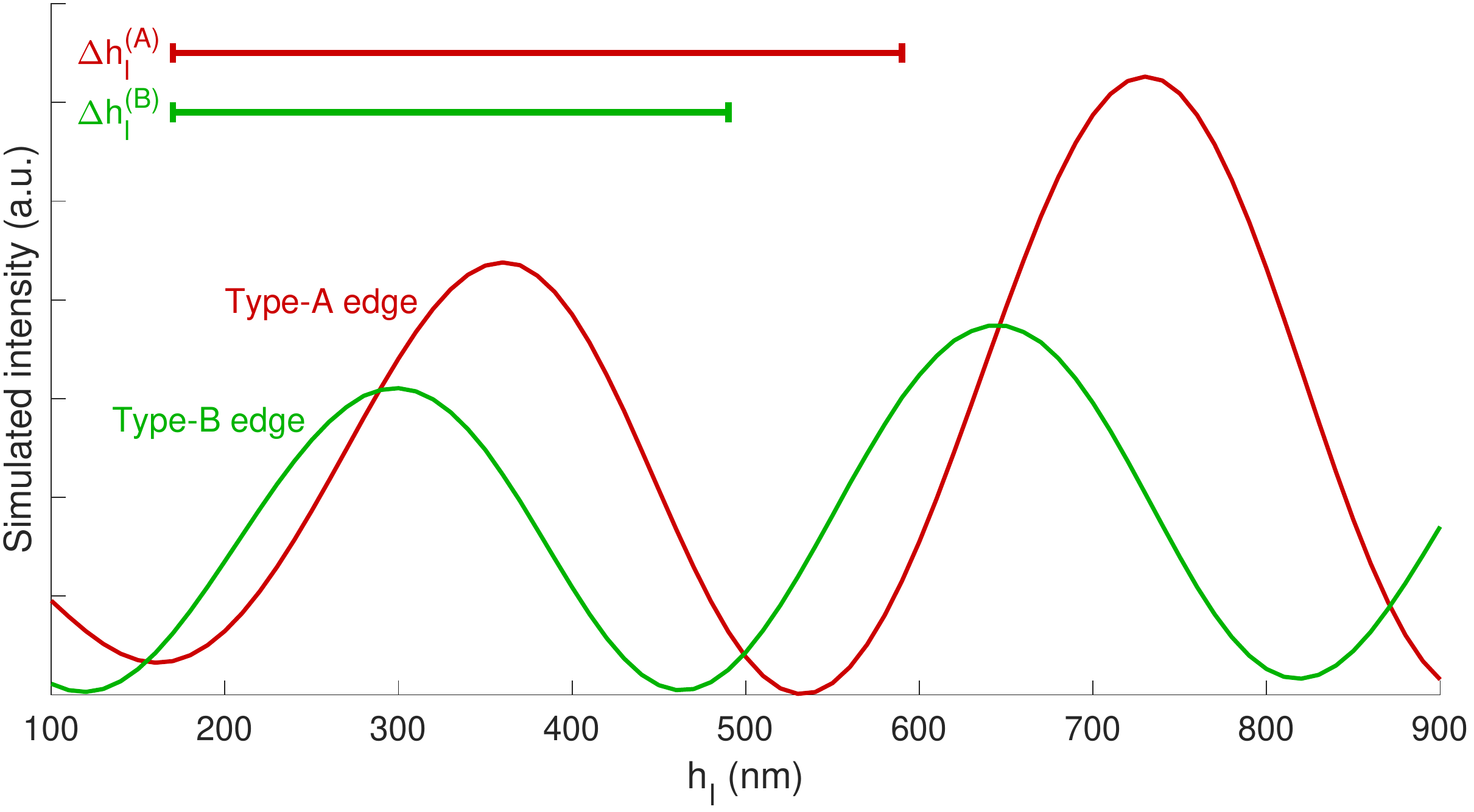}
\end{center}\linespread{1.3}
\caption{Simulated scattering intensity at $\lambda_0=700$\si{\nano \meter} for the two types of edges with $h_u=130$\,nm and range of $h_l$ values. Estimated periodicity of the oscillations are $\approx380$\,nm for type-A and $\approx340$\,nm for type-B edge. The horizontal lines show the estimates for $\Delta h_l^{(A)}\approx420$\,nm and $\Delta h_l^{(B)}\approx320$\,nm, according to Eq.~(\ref{eq:Delta}).}\label{fig:hl_oscillations}
\end{figure}

These values for the periodicity are to be compared to the numerically simulated spectra in the Figure~\ref{fig:hl_oscillations}, 
where the upwards scattered power for both types of edges at a fixed wavelength (700\,\text{nm}) is plotted as a function of the lower flake thickness $h_l$ with all other parameters (e.g. $\theta_{\text{inc}}$, $h_u$) kept as in Figure~\ref{fig:spectra_comparison}.
We observe distinct oscillations with a periodicity of $\approx 340$\,nm for type-B edges, which is in good agreement with the simple interference model. 
The periodicity of the type-A spectra is slightly less consistent as the spectrum is not a pure sinusoid (distances between minima, maxima and turning points give different "periodicities").
Anyhow, the periodicity is greater than for type-B edges and we extract a value of $\approx 380$\,nm.
This is still in qualitative agreement with the interference model.
We attribute the disparity to the existence of a second, weaker resonant effect, potentially a Fabry--P\'erot-like standing wave on the lower facet.
Yet, the main effect is clearly visible, especially since no alternative explanation predicts oscillations in the 300--400\,nm range.

So far, we have not discussed the agreement between experimental and numerical spectra (Figure~\ref{fig:spectra_comparison}).
Both panels show qualitatively similar behavior.
The most striking differences are an overall red-shift by some 100\,nm of all features in the experimental spectrum and a significantly reduced amplitude towards the blue spectral range.
This is partially due to simplifications and uncertainties in the numerical model.
Firstly, in our simulations we chose a 2D finite-element model with plane-wave illumination impinging normally on the edge.
In contrast, the illumination in our dark field experiment was from a range of azimuthal angles covering a sector of $60^\circ$.
We expect that non-normal incidence would lead to considerable red-shift of the
interference effect.
Secondly, also the elevation angle $\theta_\text{inc}$ in the experiment is not well defined.
Incident light arrives at the edge from an indefinite range of angles around approx. $12^\circ$.
In the numerical model, we assume a plane wave from $\theta_\text{inc} = 12^\circ$.
These two angular distributions lead to a smearing and potentially to a partial destruction of the interference pattern, especially at higher orders, i.e. shorter wavelength.
Thirdly, we have conducted simulations with different common models for the permittivity of gold and find that variations in the plasmon dispersion relation can easily account for 50\,nm shift in the spectral features, too. 
Although the flakes appear clean under the electron microscope, we suspect that the flake is covered by residue from the fabrication process, which again would lead to spectral red-shift and potentially to increased loss at shorter wavelengths.
Within the limits of these uncertainties, we are confident, that we have identified the main origin for the thickness-dependent dissimilar scattering spectra from type-A and type-B edges in our monocrystalline gold flakes.
Finally, the strong sensitivity with respect to the plasmon dispersion relation also offers the possibility to independently verify ellipsometrically determined material parameters of monocrystalline metal particles provided they are known to be clean and using plane-wave like illumination e.g. in a goniometer.

\section{Conclusions}

To summarize, in this work we have exemplified that the differences in the scattering spectra of the adjacent edges of the gold monocrystalline flakes, which we have first observed experimentally in the DF microscope, are far-field manifestation of the subwavelength-scale morphological features. We have developed a numerical model and filtering method which allows to simulate the experimental conditions fairly accurately. 
Through a careful analysis of numerical simulations, we found that the height of the lower trapezoid in the cross-section of the flake ($h_l$) is the main parameter for determination of the scattering spectrum. Guided by analysis of numerical results, we developed an analytic model where the physical mechanism, which gives the main contribution to the observed scattering spectrum, is the interference between a surface plasmon in the lower facet of the flake's edge and free space waves.
The difference in lengths of the facets of adjacent edges explains the difference in scattering spectra of those.

We speculate that DF spectroscopy can potentially be used for estimation of the flake thickness or the dielectric function of the monocrystalline gold. Given that the flakes have well-determined geometries, and with possibilities for experimental measurement of length scales at atomic-scale resolution, the later appears to be a more realistic task. While we might not know the exact dielectric function (tensor), we accurately know dimensions of the box containing the plasmons, which is a rare condition considering the situation where geometries are fabricated with top-down approaches such as focused ion beam (FIB)  or electron beam lithography (EBL). In other words, the accurate information about geometry may in turn be used to determine the (bulk) optical properties of the metal, by perfectly matching accurate experiments with simulations, with the dielectric function (tensor) being the unknown in the simulations.

\section{Methods}
In the following subsections experimental and numerical methods used in this work are described. 

\subsection{Sample preparation}

Gold monocrystalline flakes were prepared using the modified Brust–-Schiffrin\cite{Brust:1994:JChemSoc} method for colloidal gold synthesis in a two-phase liquid-liquid system via thermolysis \cite{Radha:2010:CrystGrowthDes}.
In this method, an aqueous solution of the chloroauric acid (\ch{HAuCl4 * 3 H2O}) in concentration \SI{0.5}{\gram\per\mole} is used as the precursor. It is mixed with a solution of tetraoctylammonium bromide (TOABr) in toluene in a vial and stirred using a magnetic stirrer for approximately 10 minutes at 5000 RPM. During this process \ch{AuCl4-} ions are transferred from aqueous solution to the toluene and TOABr acts as a phase transfer catalyst. After that the mixture is left in rest for approximately 10 minutes, during which two phases -- aqueous and organic -- separate. 
The substrate (n-type Si wafer) is prepared: a piece of silicon wafer is pre-cleaned using ultrasonic bath in acetone, isopropyl alcohol (IPA) and ultrapure water (Milli-Q). After drying with nitrogen gas the substrate is pre-baked on a hot plate at \SI{200}{\degreeCelsius} for approximately 5 minutes for dehydration purposes. In the following step few microlitres of the organic phase are drop-casted onto a substrate which is then kept on the hot-plate at \SI{160}{\degreeCelsius} for 30 minutes. After that the sample is cleaned in toluene at \SI{75}{\degreeCelsius} temperature, acetone and IPA, which removes the greater part of the organic solvent. After the sample is dried with a mild nitrogen blow, a large number and variety of gold flakes are found on the surface of the substrate.

\subsection{Spectroscopy}

DF spectroscopy measurements were performed using the Zeiss Observer microscope (Epiplan-Neofluar HD objective 50x, NA=0.80) and Andor Kymera 193i spectrograph equipped with Andor Newton CCD camera.
Additionally, two lenses (achromatic doublets with focal lengths 15 and 20\,cm) and an iris diaphragm were used to create the so-called 4f correlator system for spatial filtering (i.e. NA selection). For the measurements described in this work we have calibrated diaphragm opening to correspond to $\text{NA}\approx 0.4$. This value of the collection NA was chosen because it gives good contrast between two types of edges, while keeping sufficiently strong signal and high spatial resolution.  Lenses with different focal lengths were chosen intentionally to obtain appropriate image magnification on the camera screen.

A standard tungsten-halogen lamp was used as an illumination unit in this setup. In order to achieve "one-sided" illumination, the DF mirror cube was modified to restrict the range of azimuthal angles of incidence, i.e. DF illumination ring was partially covered with an opaque sheet and only a sector of $\approx60^\circ$ was left open. Reported spectra were normalized to a reference spectrum, obtained by illuminating a white scatterer in same conditions.

Additionally, a linear polarizer and analyzer were used to select appropriate (i.e. perpendicular to the specified edge of the flake) polarization of the incident and scattered light.

In order to avoid any systematic error due to the experimental setup, we measured different edges by rotating the sample while keeping fixed all other settings. 

\subsection{Numerical simulations}

Numerical simulations were performed using a commercially available finite-element method (FEM) solver (Comsol Multiphysics 5.3). The geometry of the model is two-dimensional, implying homogeneity along $z$-axis direction (axis orthogonal to the plane of the flake's cross-section). We consider this simplification to be appropriate as lateral dimensions of the flake are much larger than the thickness (i.e. $w_\text{fl} \gg h_u+h_l$).
The model assumes a plane wave with wavelength $\lambda_0$ incident at angle $\theta_\text{inc}$. In the first step, the model solves for the electric field distribution in the vicinity of the air/silicon interface. For the refractive index of Si, we used interpolated experimental data by Aspens\cite{Aspens:nSi:1983}. In the second step, a gold particle is placed on the substrate with the shape shown in Figure~\ref{fig:schem} and fields calculated in the first step are used as a background source to obtain the scattered fields. The refractive index of monocrystalline gold is described by interpolated experimental data from Olmon \cite{Olmon:nAu:2012}. 
The model uses triangular meshing (5\,nm maximum element size in the metal domain, 8\,nm in silicon) and fourth order polynomial basis functions. We have performed mesh refinement study, from which we assess second order convergence and estimate relative error in the reported numerical data to be less then one percent.  

In the subsequent step, the simulated fields were post processed using a dedicated filtering method, which mimics operation of the microscope objective, i.e. selects only traveling waves which propagate within a given NA.

\begin{acknowledgement}
We thank C.~Tserkezis and J.-S. Huang for stimulating discussions, and P.~Trimby (Oxford Instruments) for experimental assistance characterizing the crystal directions of the gold flakes.
C.~W. and F.~T. acknowledge funding from MULTIPLY  fellowships under the Marie Sk\l{}odowska-Curie COFUND Action (grant agreement No. 713694). S.~I.~B. acknowledges the European Research Council (grant 341054, PLAQNAP). N.~A.~M. is a VILLUM Investigator supported by VILLUM FONDEN (grant 16498). Center for Nano Optics is financially supported by the University of Southern Denmark (SDU 2020 funding). Simulations were supported by the DeIC National HPC Centre, SDU.
\end{acknowledgement}

\newpage
\bibliography{references}

\end{document}